\documentclass[aps,prl,twocolumn,amsmath,amssymb,floatfix,longbibliography,superscriptaddress]{revtex4-2}
\usepackage{color}
\usepackage{mathrsfs}
\usepackage{graphicx}
\usepackage{dcolumn}
\usepackage{bm}
\usepackage{empheq}
\usepackage[hidelinks]{hyperref}
\usepackage{xcolor}
\hypersetup{
    colorlinks,
    linkcolor={blue},
    citecolor={blue!80!black},
    urlcolor={blue!80!black}
}
\usepackage[toc,page]{appendix}
\usepackage[normalem]{ulem}
\usepackage{amsmath,amsfonts,amssymb,ulem}
\usepackage{epstopdf}
\usepackage{xcolor}
\usepackage[separate-uncertainty]{siunitx}
\usepackage{upgreek}
\usepackage{hyperref} 

\usepackage{graphicx}

\begin{document}


\title{Electron penetration heating in turbulent magnetic loops driven by nonrelativistic laser-plasma interaction}
\author{Zheng Gong}
\email[]{zgong92@itp.ac.cn}
\affiliation{Department of Mechanical Engineering, Stanford University, Stanford, California 94305, USA}
\affiliation{Institute of Theoretical Physics, Chinese Academy of Sciences, Beijing 100190, China}
\author{Sida Cao}
\affiliation{Department of Mechanical Engineering, Stanford University, Stanford, California 94305, USA}
\author{Caleb Redshaw}
\affiliation{Department of Mechanical Engineering, Stanford University, Stanford, California 94305, USA}
\author{Matthew R. Edwards}
\email[]{mredwards@stanford.edu}
\affiliation{Department of Mechanical Engineering, Stanford University, Stanford, California 94305, USA}

\date{\today}
\begin{abstract}

Using particle-in-cell simulations to study nonrelativistic laser pulse propagation in a under-critical plasma, we identify a novel mechanism that occurs during the growth of turbulent magnetic loops: electron penetration heating. The loops have an electromagnetic left-hand chirality distinct from that of well-known quasistatic magnetic islands. The fast electrons penetrate through the loops and thus are accelerated to unexpected relativistic energies due to the symmetry breaking induced by the coupling between the loop field and the non-relativistic electromagnetic wave. The identified features of penetration heating and magnetic loops might provide an alternative perspective for understanding superponderomotive electron heating in under-critical plasmas irradiated by nonrelativistic laser pulses. This is a potential explanation for anomalous hot electron generation in scenarios of laser-driven inertial confinement fusion.


\end{abstract}
\maketitle

Efficient energy coupling between nonrelativistic laser pulses and under-critical plasmas remains a cornerstone of high-energy-density physics (HEDP)~\cite{drake2006introduction}, particularly in managing the deleterious preheating effects in inertial confinement fusion (ICF)~\cite{lindl1995development,glenzer2010symmetric}. While sub-relativistic laser intensities ($a_0 \ll 1$) are conventionally expected to undergo linear propagation, the onset of long-pulse nonlinear instabilities, such as stimulated Raman scattering (SRS) and filamentation, triggers a complex energy cascade from coherent optical fields into turbulent micro-structures~\cite{kruer2019physics,montgomery2016two}. In the corona of an ICF capsule, where plasma scale lengths can reach millimeters~\cite{craxton2015direct,tabak2005review}, these instabilities can reach nonlinear saturation, leading to the production of hot electrons with energies significantly exceeding the thermal background~\cite{estabrook1980heating,riconda2016raman}. Understanding the transition from laser-driven waves to kinetic particle heating is essential for optimizing laser-target coupling in ignition-scale experiments~\cite{campbell2017laser,meezan2017indirect,scott2021shock,batani2023future,he2015physical,zhang2020enhanced}.

In recent years, a large amount of research has focused on how nonrelativistic pulses generate unexpectedly energetic electrons when propagating through under-critical plasmas~\cite{batani2019progress,tikhonchuk2019studies,yan2012generating,li2020pump,cao2020cogeneration,jiang2024anomalous,rovere2026effects}. Traditional models primarily attribute hot electron production to the wave-breaking of Langmuir waves excited by SRS~\cite{estabrook1980heating,baldis1991coexistence,rousseaux1992suprathermal}. However, recent kinetic simulations and experiments have identified that in the nonrelativistic picosecond regime, the interaction enters a strongly coupled nonlinear stage where laser filamentation instability plays a crucial role~\cite{afshar1992evidence,gu2021multi,liu2023parametric}. The laser filamentation is associated with self-focusing effects, in which nonlinear refractive index modification breaks the laser beam into multiple filaments and concentrates electromagnetic (EM) energy locally~\cite{michel2023introduction}. 
For the scenarios with a relativistically intense driving pulse ($a_0>1$), the accumulated EM wave in filamentation exerts ponderomotive expulsion on electrons to create ion density channels~\cite{pukhov1996relativistic,pukhov1999particle}, where localized electric fields can facilitate the generation of superponderomotive electrons via suppressing the electron dephasing in laser EM fields~\cite{robinson2013generating,arefiev2016beyond}. Furthermore, the evolution of these filaments can also lead to the formation of post-solitons to trap EM energy in density cavities~\cite{esirkepov1998low,bulanov1999solitonlike,naumova2001formation,borghesi2002macroscopic,shen2004ultrashort,lezhnin2018annihilation}. These structures could persist long after the pulse passage, potentially acting as sites for localized energy dissipation and electron heating through stochastic or resonant interactions in the nonrelativistic laser-plasma interaction~\cite{gu2021multi,liu2023parametric}, but these are rarely investigated.

In this work, we employ particle-in-cell (PIC) simulations to investigate the generation of superpondermotive electrons in nonrelativistic picosecond-scale laser-plasma interactions.
We identify a novel mechanism---electron penetration heating---arising during the evolution of turbulent magnetic loops that exhibit an EM left-hand chirality. As electrons penetrate these magnetic loops, an additional transverse deflection disrupts the symmetry of their energy exchange with the EM wave, enabling acceleration to unexpected MeV energies despite the non-relativistic intensity of the driving wave. 
This mechanism can potentially bridge the gap between macro-scale laser propagation and micro-scale kinetic heating, providing an efficient pathway for the observed MeV-scale populations in large-scale laser experiments~\cite{batani2019progress}.

The simulations were performed with the PIC code EPOCH~\cite{arber2015contemporary} using parameters relevant to the laser experiment, although the results are straightforwardly scaled to astrophysical parameters.
The $100\mathrm{\mu m}\times 40\mathrm{\mu m}$ two-dimensional (2D) simulation domain was captured on a $4000\times 1600$ grid. 
A linearly $s$-polarized laser pulse was incident from the left boundary.
Our main example used a peak intensity of $I_0=5.5 \times 10^{16}\mathrm{W/cm}^2$, equivalent to the normalized amplitude $a_0\equiv eE_l/(m_ec\omega_0)= 0.2$ with a wavelength $\lambda_0=2\pi c/\omega_0=1\mu m$ and $c$ the speed of light. 
The pulse was transversely infinite and had a duration of $4\,$ps. 
The plasma electron and proton densities were $n_e=n_i=0.05n_c\approx 6\times10^{19}\mathrm{cm}^{-3}$, and the proton mass is $m_i=1836m_e$. The temperature is $T_{e,i}=10\,$eV for both species.
A periodic boundary condition was used for the lateral sides and an open condition was used for the longitudinal boundaries. Other detailed parameter setup in PIC simulations can be found in Table~\ref{tbl:params} below.
Assuming a laser pulse with a transverse spot size of $300\,\mu m$, the corresponding power of $156\,\text{TW}$ and energy of $621\,\text{J}$ are consistent with the state-of-the-art capabilities of long-pulse laser-plasma interaction facilities~\cite{stuart2006titan,maywar2008omega,danson2015petawatt_review}.

\begin{widetext}

\begin{table}[h]
\caption{Physical and Computational Parameters for the electron penetration acceleration in turbulent magnetic loops\\ \\}
\label{tbl:params}
\begin{ruledtabular}
\begin{tabular}{l c c c c c c}
\noalign{\smallskip}
{\bf Figure} &
Cells/$\lambda_0$\footnote{All dimensional values are calculated using $\lambda_0 = 1\ \mathrm{\mu m}$, but with the neglect of atomic physics and appropriate non-dimensionalization the simulations results are not $\lambda_0$-dependent.} &
Part./Cell\footnote{Number of particles in a cell in regions where plasma density is non-zero at the beginning of the simulation.}&
$a_0$ &
$\tau/T_0$\footnote{All pulses have a trapezoid time profile with $\tau$ the total duration and the up and down ramp part are 5\% of the whole duration, respectively.} &
$n_e/n_c$\footnote{$n_e$ is the electron number density and $n_c$ is the critical density for light at $\lambda_0$ ($n_c =  m_e \omega_0^2 / 4 \pi e^2$ with $\omega_0=2\pi c/\lambda_0$).} &
$m_i/m_p$ \\
\noalign{\smallskip}
\hline
\noalign{\medskip}
1,2,3       & 40 & 40 & 0.2  & 1200 & 0.05 &  1 \\
4(a)(c) & 40 & 40 & 0.2  & 1200 & 0.05 &  1 \\
5(a)(d) & 40 & 40 & 0.2  & 1200 & 0.05 &  1  \\
6        & 40 & 40 & 0.2  & 1200 & 0.05 &  1 \\
7(a)    & 40 & 40 & 0.05, 0.2, 0.4  & 4800 & 0.01--0.4 &  1 \\
7(b)(c) & 40 & 40 & 0.01--0.4  & 4800 & 0.0005--0.4 &  1 \\
7(d)    & 40 & 40 & 0.01--0.4  & 4800 & 0.01--0.4 &  1 
\end{tabular}
\end{ruledtabular}
\end{table}

\end{widetext}

\begin{figure*}
\centering
\includegraphics[width=0.85\textwidth]{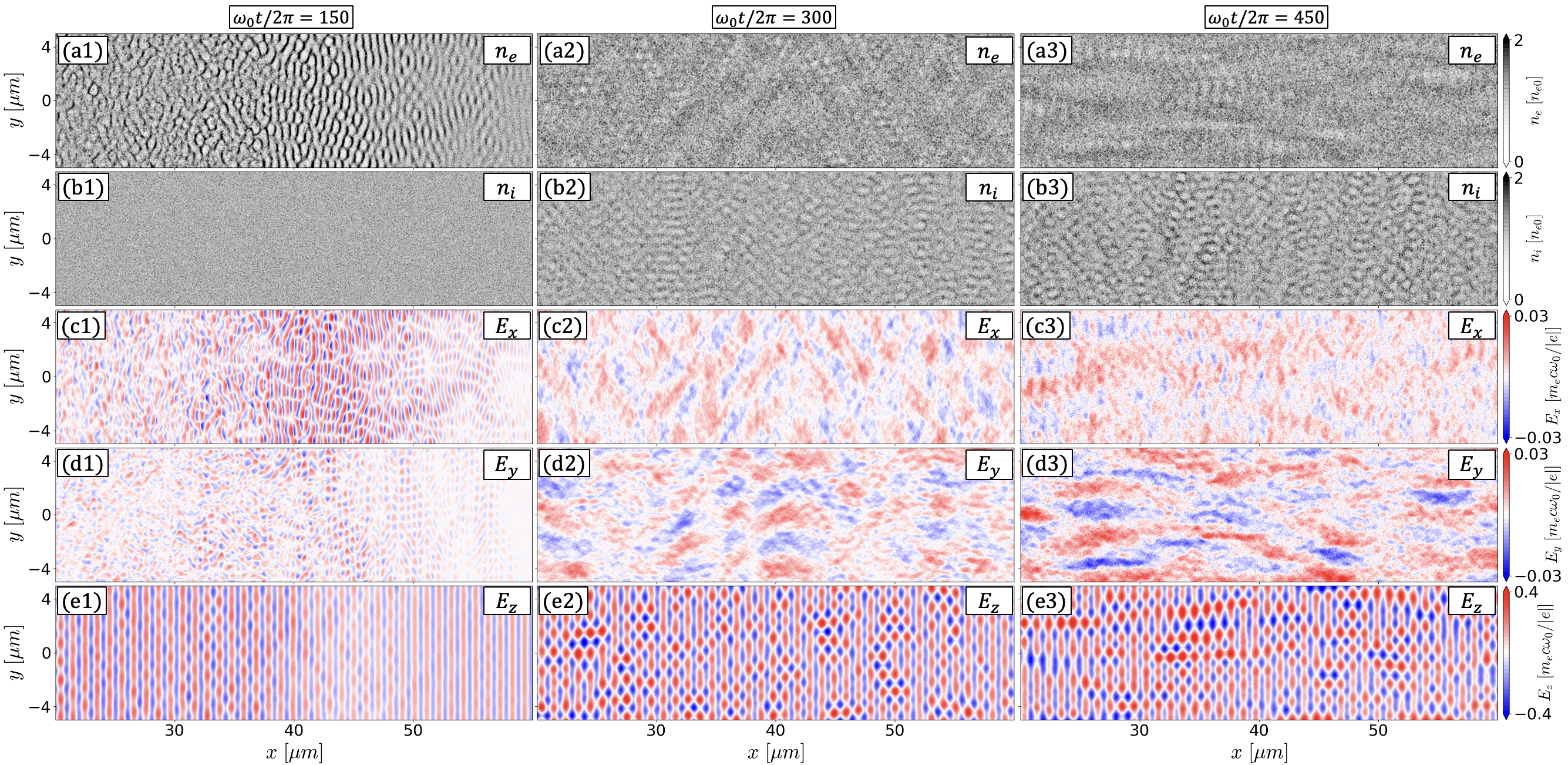}
\caption{PIC simulation results. The spatial distribution of (a) electron density $n_e$, (b) ion density $n_i$, (c) longitudinal electric field $E_x$, (d) transverse electric field $E_y$, and (e) EM field $E_z$. Here, the three columns present the results at the time of $\omega_0t/2\pi = 150$, $300$, and $450$.}
\label{fig:res_den_Exyz}
\end{figure*}

\begin{figure*}
\centering
\includegraphics[width=0.65\textwidth]{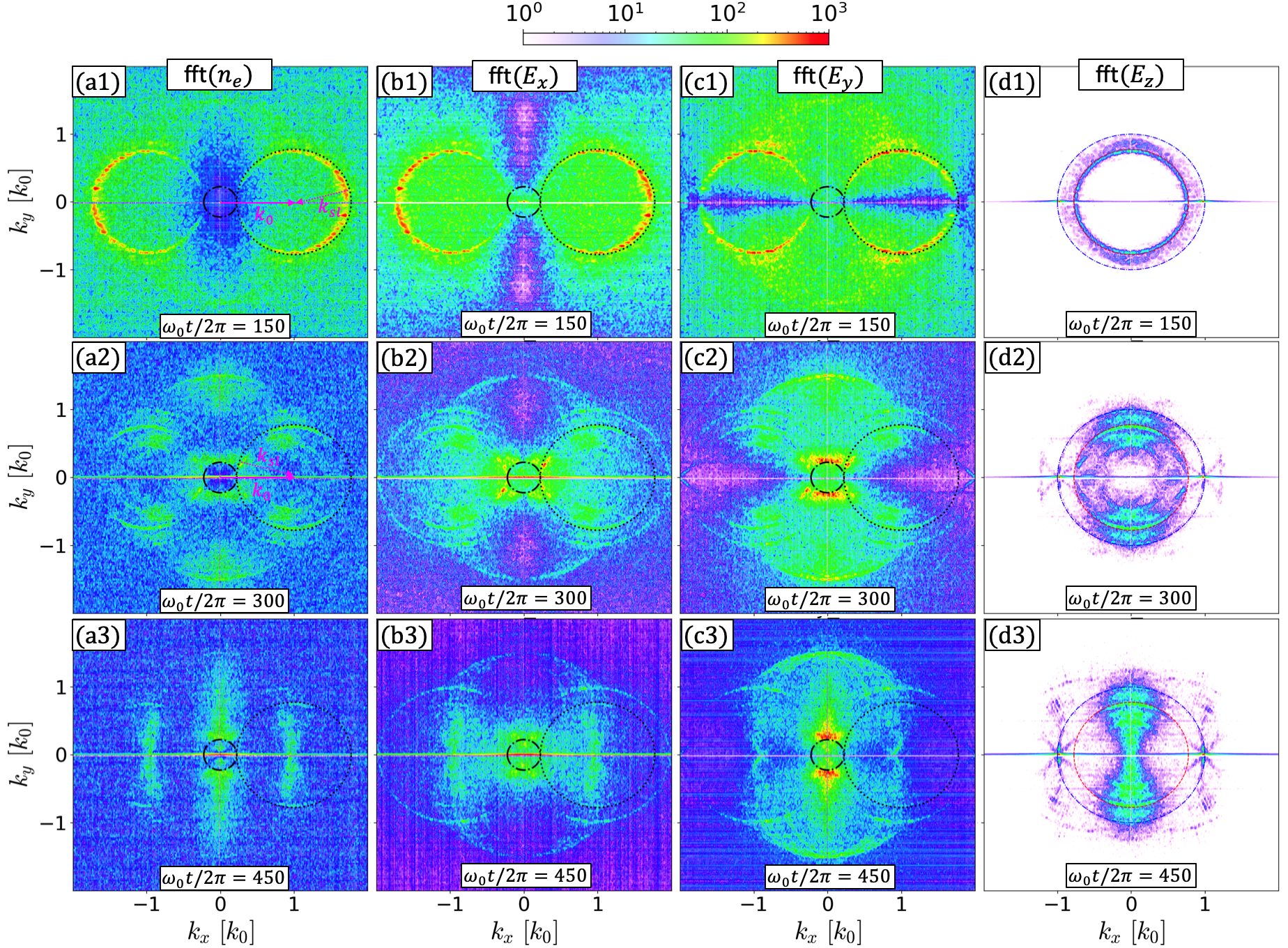}
\caption{PIC simulation results. The fast Fourier transform (fft) spectra for (a) electron density $\mathrm{fft}(n_e)$, (b) electric field $\mathrm{fft}(E_x)$, (c) electric field $\mathrm{fft}(E_y)$, and (d) EM field $\mathrm{fft}(E_z)$. Here, the three rows present the results at the time of $\omega_0t/2\pi = 150$, $300$, and $450$. In (a)(b)(c), the black dash-dotted lines refer to $(k_x^2+k_y^2)^{1/2}=k_{pe}$ while the black dotted lines represent $[(k_x-k_0)^2+k_y^2]^{1/2}=k_{sl}$. In (d), the blue dash-dotted lines refer to $(k_x^2+k_y^2)^{1/2}=k_{0}$ while the red dashed lines represent $(k_x^2+k_y^2)^{1/2}=k_{sl}$. In (a1)(a2), the magenta arrows represent the wave vectors of the pumple light and the scattered light.}
\label{fig:res_fft}
\end{figure*}

The spatial distributions of the electron density $n_e$, ion density $n_i$, longitudinal electric field $E_x$, transverse electric field $E_y$, and electromagnetic (EM) field $E_z$ at $\omega_0 t/2\pi = 150, 300, \text{ and } 450$ are presented in Fig.~\ref{fig:res_den_Exyz}. The corresponding Fast Fourier Transform (FFT) spectra are provided in Fig.~\ref{fig:res_fft}.
Our analysis shows that $E_x$ and $E_y$ are primarily sustained by charge separation between locally accumulated electron populations and the slower-responding ions. Notably, comparative simulations using immobile ions yielded results similar to the mobile-ion case, confirming that the initial electron dynamics are the primary driver of these fields.

At the beginning $\omega_0t/2\pi=150$, the electron density $n_e$ exhibits a modulation with a scale length comparable to the laser wavelength $\delta l_s\sim 2\pi c/\omega_0 =1\mu m$. Consequently, the quasi-static fields $E_x$ and $E_y$ exhibit similar modulated structures. The modulation on the spatial distribution of EM field $E_z$ is also observable. The FFT spectrum of $E_z$ displays a ring structure at $|\bm{k}|=k_{sl}$ [see red dashed line in Fig.~\ref{fig:res_fft}(d)], where $k_{sl}=k_0-k_{pe}$ is the wavenumber of the scattered light and $k_{pe}\equiv\omega_{pe}/c$ corresponds to the cold plasma Langmuir wave. This indicates the onset of the SRS parametric instability, where the incident laser wave $k_{0}$ decays into a scattered EM wave ($k_{sl}$) and a Langmuir wave ($k_{pe}$). The electron density $n_e$ modulation results from the nonlinear coupling effect between the incident light ($\bm{k}_0$) and scattered light ($\bm{k}_{sl}$), confirmed by the ring structure at $[(k_x-k_0)^2+k_y^2]^{1/2}=k_{sl}$ (i.e. $\bm{k}=\bm{k}_0+\bm{k}_{sl}$) in the FFT spectrum of $n_e$ [see black dotted line in Fig.~\ref{fig:res_fft}(a1)]. The quasi-static electric fields $E_x$ and $E_y$ are induced by the electron density modulation and thus they have a similar distribution pattern as that of $\mathrm{fft}(n_e)$.

By $\omega_0t/2\pi=300$, the electron density $n_e$ modulation is dominated by a large scale length approximating the plasma wavelength, $\delta l_{pe}\sim 2\pi c/\omega_{pe}\approx4.5\mu m$ with $\omega_{pe}\sim (4\pi n_e e^2/m_e)^{1/2}$ [Fig.~\ref{fig:res_den_Exyz}(a2)], though the small-scale modulations persist. As determined by the electron density modulation, the electric fields $E_x$ and $E_y$ show the similar modulation at that of $n_e$, i.e. with a large scale size of $\delta l_{pe}$ [Fig.~\ref{fig:res_den_Exyz}(c2)(d2)]. The corresponding FFT spectra indicate that the plasma Langmuir wave becomes prominent [Fig.~\ref{fig:res_fft}(a2)(b2)(c2)], and the scattered light direction begins to shift from backward to lateral side-scattering [see Fig.~\ref{fig:res_fft}(a2)(b2)(c2)].
As shown in Fig.~\ref{fig:res_fft}(d2), a distinct concentric structure is observed passing through $k_y \approx \pm (k_0 - 2k_{pe})$. This suggests that the Langmuir waves (characterized by $E_y$ and $k_y \approx \pm k_{pe}$) can be interpreted as a secondary scattering of the already side-scattered EM waves.
At $\omega_0t/2\pi=300$, the electron density exhibits a pronounced transverse filamentation distribution with a scale length of $\delta l_{pe} \sim 2\pi c/\omega_{pe}$ [Fig.~\ref{fig:res_den_Exyz}(a3)]. In the quasi-static field distribution, $E_y$ becomes the dominant component while $E_x$ remains secondary [Fig.~\ref{fig:res_den_Exyz}(c3)(d3)]. The FFT of $\mathrm{fft}(E_z)$ [Fig.~\ref{fig:res_fft}(d3)] and the spatial modulation of $E_z$ [Fig.~\ref{fig:res_den_Exyz}(e3)] clearly demonstrate the dominance of side-SRS.

\begin{figure*}
\centering
\includegraphics[width=0.85\textwidth]{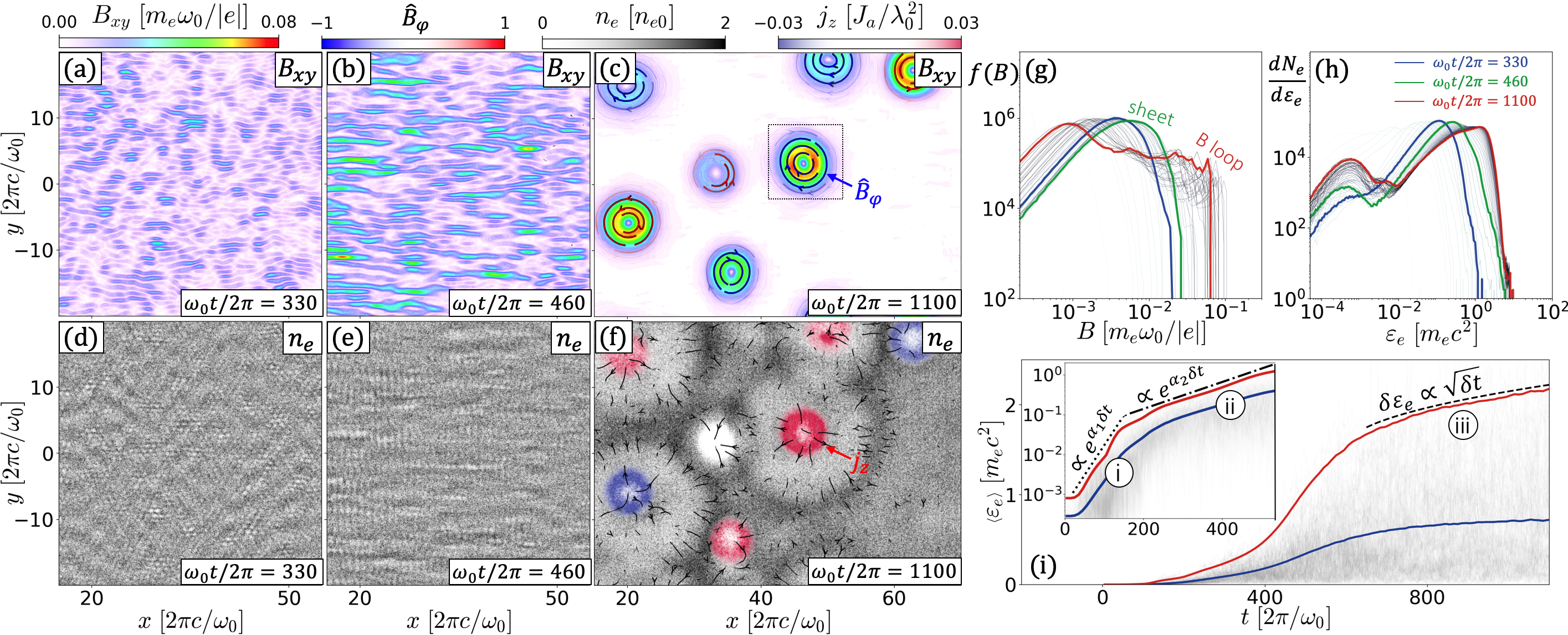}
\caption{PIC simulation results. (a)-(c) and (d)-(f) present the distribution of magnetic field $B_{xy}\equiv(B_x^2+B_y^2)^{1/2}$ and electron density $n_e$ at $\omega_0 t/(2\pi)$=330,\,460,\,and\,1100, respectively~\cite{animation_fig3_bxy_ne}. In (c), the blue-red arrows denote the direction of magnetic loops $\hat{B}_\varphi$ with $B_{\varphi}\equiv -B_x\sin(\varphi)+B_y\cos(\varphi)$. In (f), the blue-red color shows current density $j_z$ while black arrows represent the electric field direction.
(g) and (h) show the time-evolved magnetic spectra $f(B)$ vs $B$ and electron energy spectra $dN_e/d\varepsilon_e$ vs $\varepsilon_e$, respectively. (i) Time evolution of the electron energy $\left<\varepsilon_e\right>$ vs $t$, where blue (red) line corresponds to the average over all (top 10\% energetic) electrons while the black lines denote the fitted scaling laws. The results in (g)(h)(i) are calculated over the entire simulation domain.}
\label{fig:fig2_general}
\end{figure*}

\begin{figure*}
\centering
\includegraphics[width=0.75\textwidth]{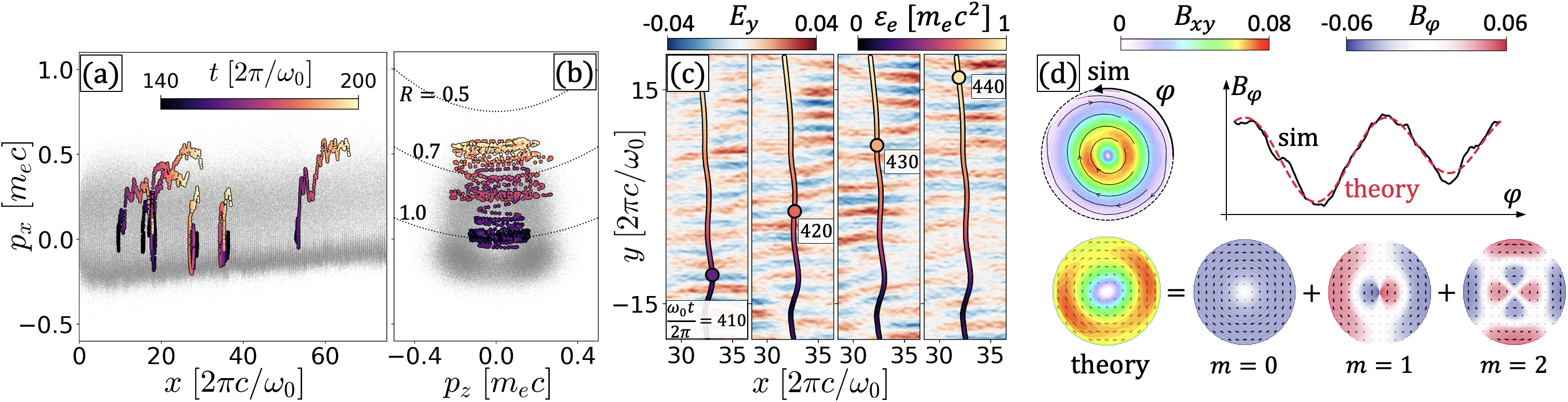}
\caption{Electron distribution in (a) $(x,p_x)$ and (b) $(p_z,p_x)$ space, where the color denotes the time evolution of typical electron trajectories. 
(c) Electron drifting acceleration in the transverse electric field $E_y$ of plasma filamentations.
(d) Comparison between the simulated and analytical magnetic loop field.}
\label{fig:fig3_b_loop}
\end{figure*}

During the interaction, the turbulent magnetic sheets gradually elongate and transform to the loop shape [see Fig.~\ref{fig:fig2_general}(a)-(c)], 
while the electron density $n_e$ transitions from oblique to filament structures and exhibits crater-like modulation [Fig.~\ref{fig:fig2_general}(d)-(f)].
In contrast to well-studied magnetic islands in current filamentation instabilities~\cite{medvedev1999generation,honda2000collective,califano2001fast,silva2003interpenetrating,califano2006three,bret2008exact}, our new feature is that the correlation between the magnetic rotation direction $\hat{B}_\varphi$ and the current density $j_z$ exhibits a left-hand chirality with $(\nabla\times\hat{B}_\varphi)\cdot j_z<0$.
This feature is because the magnetic loop originates from the EM eigenstate of the plasma crater, as explained below.
The time-evolved magnetic spectra $f(B)$ in Fig.~\ref{fig:fig2_general}(g) show that the distribution of the sheet strength is quasi-monoenergetic at $B\sim 10^{-2}m_e\omega_0/e$, but a double-bump distribution appears with the loop strength at $B\sim 10^{-1}m_e\omega_0/e$ and the weaker sheet at $B\sim10^{-3}m_e\omega_0/e$. 
The spectra $dN_e/d\varepsilon_e$ in Fig.~\ref{fig:fig2_general}(h) show electrons accelerated with a peak energy beyond MeV with the magnetic loop occurrence at $\omega_0 t/(2\pi)\sim 1100$. Here, around $2.8\%$ of the EM-driver energy is dissipated during the propagation process.
The evolution of the averaged electron energy $\left<\varepsilon_e\right>$ in Fig.~\ref{fig:fig2_general}(i) suggests three distinct interaction stages at (i) $\omega_0 t/(2\pi)<200$, (ii) $200<\omega_0 t/(2\pi)<500$, and (iii) $700<\omega_0 t/(2\pi)$, respectively.
%

In stage (i) $\omega_0 t/(2\pi)<200$, the electron energy exhibits an exponential growth $\left<\varepsilon_e\right>\propto \exp{(\alpha_1 t)}$ up to $0.1 m_ec^2$ with $\alpha_1\approx 0.005\omega_0$, one order of magnitude higher than the maximum energy $\varepsilon_e^m\sim a_0^2m_ec^2/(2R)\sim0.02m_ec^2$ predicted by an electron oscillation in a plane EM wave with $a_0= 0.2$ and a static initial condition $R= 1$~\cite{gibbon2005short}. The electron dephasing value $R\equiv \gamma_e-p_x/(m_ec)$ can be reduced by stimulated plasma fields~\cite{meyer1999electron}, which leads to the formation of forward moving electrons. Meanwhile, the backward electron flow tends to compensate the current density to sustain a counter-streaming flow [see Fig.~\ref{fig:fig3_b_loop}(a)].
In stage (ii) $200<\omega_0 t/(2\pi)<500$, 
the scattered light direction begins to shift from backward to lateral (side-scattering).
After the transition, the filamentary electron density distribution results in a quasi-static electric field $E_y$, in which electrons can stay in a favorable phase and get accelerated to an energy of $\varepsilon_e\sim m_ec^2$ [see Fig.~\ref{fig:fig3_b_loop}(c)]. This acceleration accounts for the energy enhancement $\left<\varepsilon_e\right>\propto \exp{(\alpha_2 t)}$ with $\alpha_2=0.001\omega_0$ at $\omega_0 t/2\pi\sim400$ [Fig.~\ref{fig:fig2_general}(i)].

\textit{Turbulent magnetic loops}---
At the final stage (iii) $700<\omega_0 t/(2\pi)$, when the transverse filamentation appears, the density perturbation $\delta n$ imprints an inhomogeneous refractive index $N=[1-(n_{0}+\delta n)/n_c]^{1/2}$. Based on Fermat's principle with $\partial N/\partial y\neq 0$, the EM wave tends to be focused at the density valley. As the ponderomotive force $F_p\propto - \nabla E_z^2$ of the focused EM wave can further expel the electrons out of the density valley with $\delta n<0$~\cite{kaw1973filamentation,max1974self,antonsen1992self,michel2023introduction}, the density cavities are induced [see Fig.~\ref{fig:fig2_general}(f)].
The developed cavities would convert the propagating EM wave to the eigenstate $E_z(r,\varphi,t)$ characterized by $\mathcal{R}^2(d^2 g/d\mathcal{R}^2)+\mathcal{R}(dg/d\mathcal{R})+(\mathcal{R}^2-m^2)g=0$~[see Appendix~B], where $E_z=E_mg(r)e^{im\varphi}e^{i\omega_b t}$, $\mathcal{R}=(\omega_b/c)r$, and $\varphi=\mathrm{atan2}(y,x)$. Here, $\omega_b$ is the eigenfrequency of the cavity and the integer $m=0,1,2,3...$ denotes the azimuthal mode.
The solutions of $g(\mathcal{R})$ are Bessel functions $ J_m(\mathcal{R})$ and the general eigenstate reads $ E_{z,m} = E_m J_m(k_b r) e^{i m\varphi} e^{i\omega_b t}$, where $ k_b=\omega_b/c$ is determined by $J_m(k_b r_b) = 0$ at the edge $r=r_b$.
The corresponding $B_{\varphi,m}$ is derived via $\partial B_{\varphi,m}/\partial t=\partial E_{z,m}/\partial r$ as
\begin{equation}\label{eq:B_loop}
   B_{\varphi,m} = B_m J_m'(k_b r) e^{i m \varphi} e^{i\omega_b t},
\end{equation}
where $B_m=-i(k_b/\omega_b)E_m$.
Figure~\ref{fig:fig3_b_loop}(d), showing the dotted box in Fig.~\ref{fig:fig2_general}(c), demonstrates that the magnetic loop can be decomposed into three main components $B_\varphi \approx \sum_{m=0}^2 B_{m}J_m'(k_cr) e^{im\varphi}e^{i\omega_b t}$ with amplitudes $B_{0}\approx 0.06$, $B_{1}\approx 0.001$, and $B_{2}\approx 0.003$, which verifies the eigenstate feature of the turbulent magnetic loops. 
Given the vector potential of the loop field $A_z$, $E_z=-\partial A_z/\partial t=-i\omega_bA_z$, $\nabla\times \hat{B}_\varphi\propto \partial E_z/\partial t=\omega_b^2A_z$ and $p_z\sim A_z$, the current density is estimated as $j_z\sim -en_e p_z/\gamma_e\propto -A_z$ and thus $(\nabla\times \hat{B}_\varphi)\cdot j_z\propto -\omega_b^2A_z^2<0$. This indicates that the left-hand chirality of magnetic loops is due to the causal exchange between the current density $j_z$ and the azimuthal field $B_\varphi$ distinct from the right-hand one. The latter is common in the magnetic islands of current filamentation instabilities~\cite{weibel1959spontaneously,lee1973electromagnetic,medvedev1999generation,honda2000collective,califano2001fast,silva2003interpenetrating,califano2006three,bret2008exact,alves2012large,Grassi2017_weibel,li2021nanoscale,gong2023electron} and kinetic turbulence reconnection~\cite{hoshino2012stochastic,matsumoto2015stochastic,comisso2018particle,comisso2021pitch} with $(\nabla\times\hat{B}_\varphi)\cdot j_z>0$ predicted by $\nabla\times \bm{B}=4\pi \bm{j}/c$.

\begin{figure}
\centering
\includegraphics[width=0.48\textwidth]{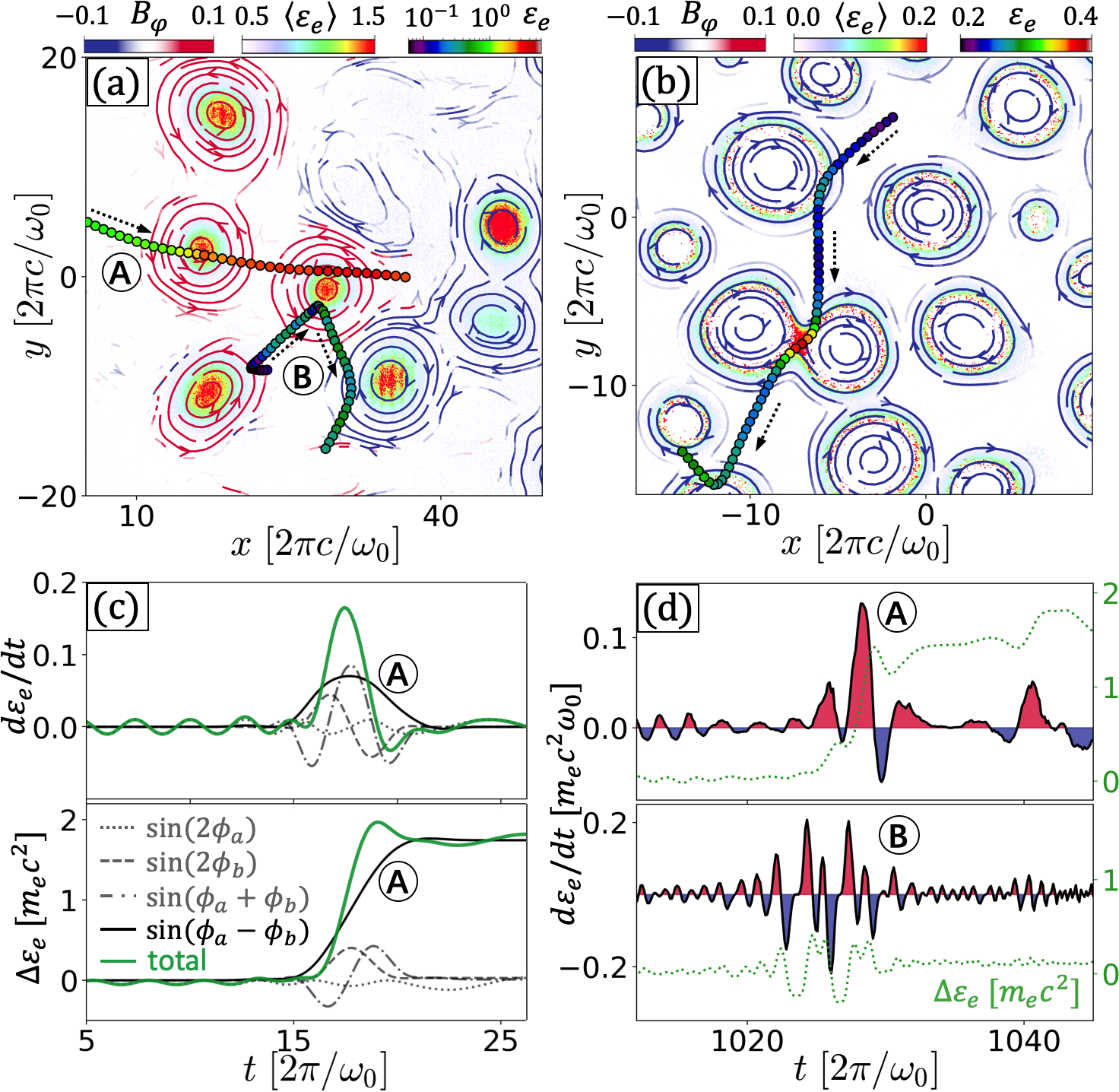}
\caption{(a) Electron penetration acceleration in magnetic loops~\cite{animation_fig5a_acc_loop}, where blue-red color denotes the rotating direction of magnetic fields $B_\varphi$, background rainbow presents the spatially averaged electron energy $\left<\varepsilon_e\right>$, and the rainbow dots present the instantaneous energy $\varepsilon_e$ of electron `A' and `B'. 
(b) The result of a representative interaction between an electron and magnetic islands in current filamentation instabilities, see Ref.~\cite{animation_fig5b_acc_CFI} for details.
(c) Analytical~\cite{animation_fig5c_theory} and (d) simulation results of the time dependence of electric work $d\varepsilon_e/dt\approx -\bm{\beta}\cdot\bm{E}$ and energy increment $\Delta\varepsilon_e$. 
}
\label{fig:fig4_e_acc}
\end{figure}

\begin{figure}
\centering
\includegraphics[width=0.48\textwidth]{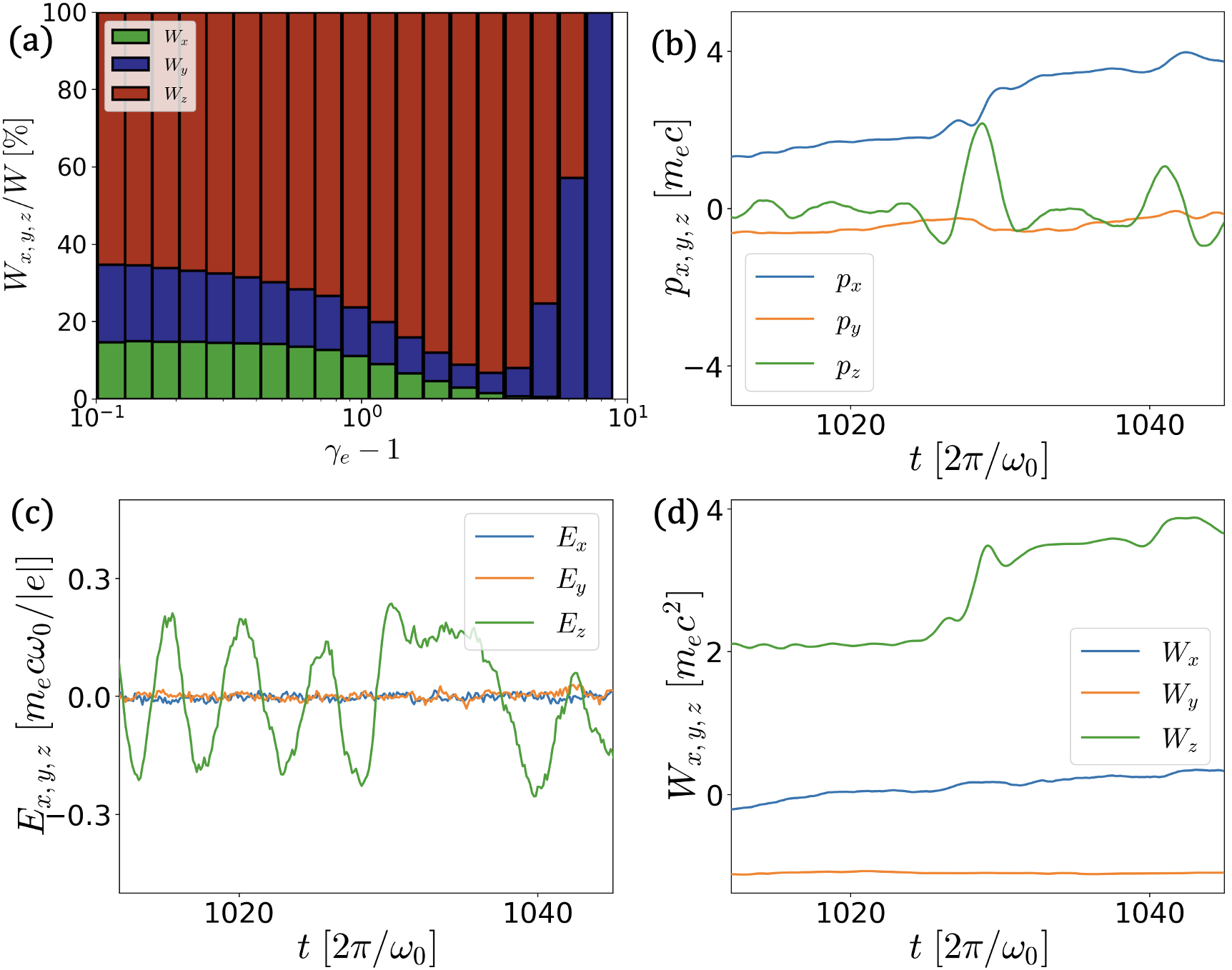}
\caption{(a) The ratio of electrons' work contribution $W_x/W$, $W_y/W$, and $W_z/W$ by the electric fields of $E_x$, $E_y$, and $E_z$ at $\omega_0 t/2\pi=1100$, respectively. $W\equiv W_x+W_y+W_z$. The other three panels show the time evolution of (b) electron momentum $p_{x,y,z}$, imposed electric field $E_{x,y,z}$, and work contribution $W_{x,y,z}$.
}
\label{fig:res_workz_pz}
\end{figure}

\textit{Penetration acceleration}---After magnetic loops emerge at $\omega_0 t/2\pi>700$, electrons are further accelerated while penetrating through the loop structure [see Fig.~\ref{fig:fig4_e_acc}(a)], which is distinct from Fermi-like stochastic acceleration during the coalescence of magnetic islands in current filamentation instabilities [see Fig.~\ref{fig:fig4_e_acc}(b)]~\cite{hoshino2012stochastic,matsumoto2015stochastic,gong2023electron}. The vector potential of the laser field is $A_{a,z}=i (E_a/\omega_0)e^{i\phi_a}$ with $\phi_a=\omega_0t-k_0x+\phi_{a0}$, and that of the magnetic loop is $A_{b,z}=i(E_b/\omega_b)J_0(k_b r)e^{i\phi_b}$ with $\phi_b=\omega_bt+\phi_{b0}$, where only the first azimuthal mode $m=0$ in $A_{b,z}$ is considered as suggested by PIC simulations. The electron momentum along the $z$-axis is derived as $p_{z}=A_{a,z}+A_{b,z}$ and the energy gain is mainly contributed by $d\varepsilon_e/dt\sim -ep_zE_z/\gamma_e$ [see Appendix C] reformulated as 
\begin{small}
\begin{eqnarray}\label{eq:dgg_dt}
 \frac{d\varepsilon_e}{dt} &&\sim \frac{E_a^2}{2\omega_0}\sin(2\phi_a) + \frac{E_b^2J_0^2(k_br)}{2\omega_b}\sin(2\phi_b) + \frac{E_aE_bJ_0(k_br)}{2\omega_0\omega_b}[ \nonumber\\
&&(\omega_0+\omega_b)\sin(\phi_a+\phi_b) + (\omega_b-\omega_0)\sin(\phi_a-\phi_b) ].
\end{eqnarray}
\end{small} 
The numerical integral of Eq.\eqref{eq:dgg_dt} plotted in Fig.~\ref{fig:fig4_e_acc}(c) shows that the energy gain comes from the coupled term with $\sin(\phi_a-\phi_b)$ when the electron `A' gets deflected by magnetic loops to suppress its dephasing in the laser field. 
The time-dependent $d\varepsilon_e/dt$ and $\Delta\varepsilon_e$ in Fig.~\ref{fig:fig4_e_acc}(a)(c)(d) demonstrate that the symmetry of electron energy exchange with laser fields is broken in the penetration process and thus a pronounced energy gain $\Delta\varepsilon_e$ is accumulated for the electron `A'.
Note that penetration acceleration is exclusively effective for energetic electrons with momentum $p\gtrsim 2\pi ceB_\varphi/\omega_{pe}$, because an `B' electron with a gyroradius $r_g\sim p/(eB_\varphi)$ much less than the loop radius $r_b\sim 2\pi c/\omega_{pe}$ cannot penetrate the loop while undergoing multiple rebounds between different loops' outer edges [see Fig.~\ref{fig:fig4_e_acc}(a)(d)].
In contrast, during the stochastic acceleration process [see Fig.~\ref{fig:fig4_e_acc}(b)]~\cite{animation_fig5b_acc_CFI}, electrons undergo multiple rebounds by each magnetic filament edge; the net energy gain comes from the asymmetric electric field $E_{xy}$ at the filaments' outer sheath. A feature that distinguishes these two mechanisms is that the energetic electrons are located at edges of magnetic islands for stochastic acceleration but are located at the loop center for penetration acceleration [see $\left<\varepsilon_e\right>$ in Fig.~\ref{fig:fig4_e_acc}(a)(b)]. 
Apart from that, the spatial distribution of averaged electron energy $\left<\varepsilon_e\right>$ in Fig.~\ref{fig:fig4_e_acc}(a) illustrates that the most energetic electrons are predominantly from the penetration acceleration process. The other potential electron heating mechanism, such as the stimulated Raman scattering instability during laser propagation via excited plasma waves, is not the critical factor in our scenario.

To further confirm the electron energy gain is mainly contributed by the work $W_z$ done by the $E_z$ along the $z$ direction. The simulation results of the work contribution of $W_{x,y,z}$ and the time evolution of the electron momentum $p_{x,y,z}$, the imposed electric field $E_{x,y,z}$, and the work contribution $W_{x,y,z}$ are shown in Fig.~\ref{fig:res_workz_pz}. Here, $W_{x}\equiv -|e|\int v_xE_xdt$, $W_{y}\equiv -|e|\int v_yE_ydt$, and $W_{z}\equiv -|e|\int v_zE_zdt$ represent the work done by the electric field along the three directions.
The results of electron shown in Fig.~\ref{fig:res_workz_pz}(b)(c)(d) corresponds the `A' electron undergoing penetration acceleration at $\omega_0t/2\pi \approx 1030$ shown in Fig~\ref{fig:fig4_e_acc}.

\begin{figure}
\centering
\includegraphics[width=0.48\textwidth]{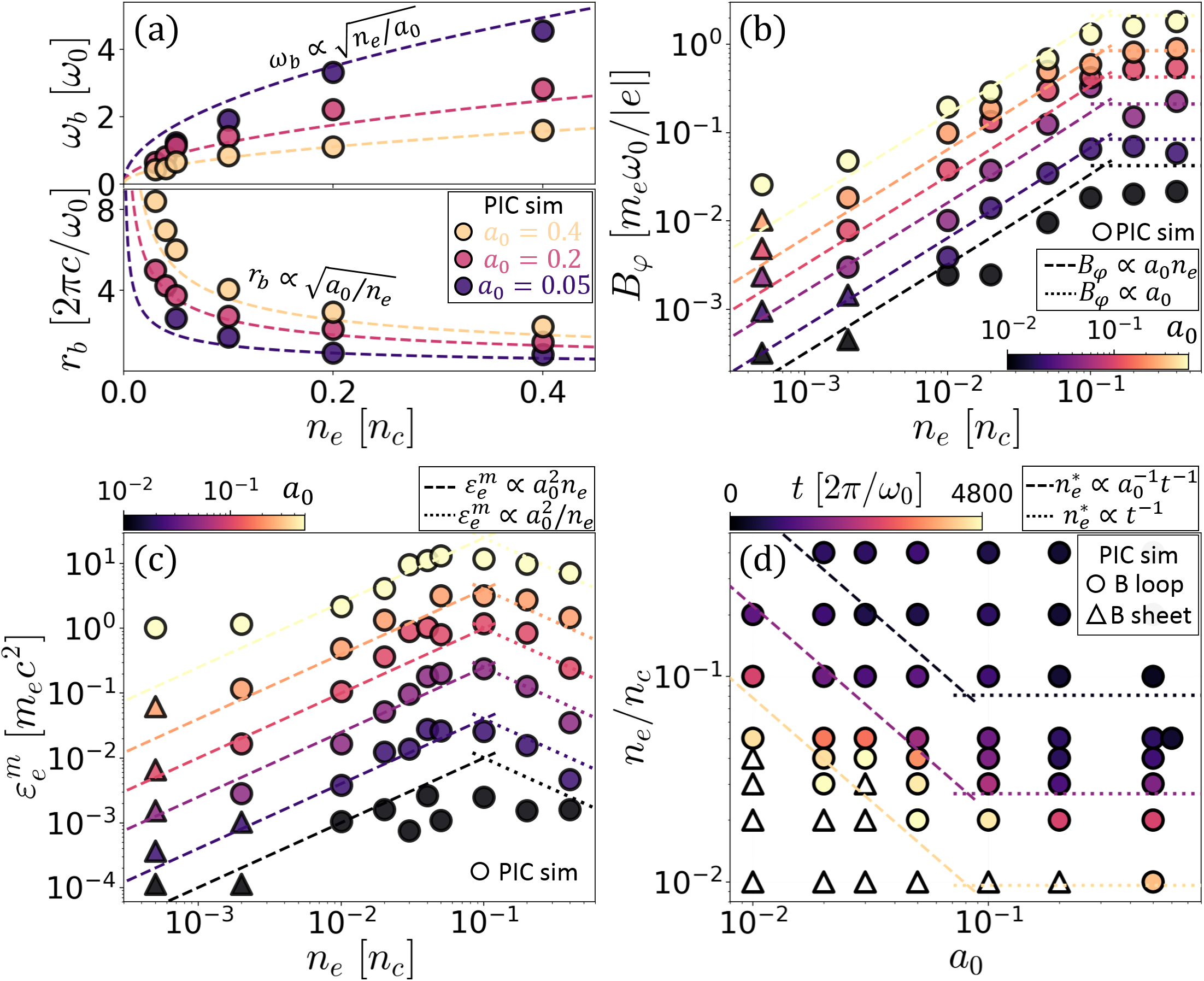}
\caption{Parameter scans of PIC simulations.
(a) The oscillating frequency $\omega_b$ and scale size $r_b$ of magnetic loops versus $n_e$. 
The dependence of (b) magnetic strength $B_\varphi$ and (c) top 10\% electron energy $\varepsilon_e^{m}$ on $n_e$ and $a_0$.
(d) Time of the magnetic loop appearance $t_b$ versus $n_e$ and $a_0$. 
Triangles denote the case without the loop appearance at $\omega_0t/2\pi\leqslant 4800$ while the lines in each panels refer to the analytical prediction.}
\label{fig:fig5_para_scan}
\end{figure}

The scale length $r_b$ and frequency $\omega_b$ of the magnetic loops are calculated through the plasma oscillating period as $r_b\sim 2\pi v_0/\omega_{pe}\sim [\pi a_0m_ec^2/(n_ee^2)]^{1/2}$ with $\omega_{pe}\sim (4\pi n_e e^2/m_e)^{1/2}$ and $\omega_b\sim j_{0,1}c/r_b\sim j_{0,1}[n_e e^2/(\pi a_0m_e^2)]^{1/2}$ [see Fig.~\ref{fig:fig5_para_scan}(a)], where $j_{m,l}$ is the $l$-th zero of $J_m(z)$, $j_{0,1}\approx2.4$, and $v_0\propto a_0^{1/2}$ fitted from simulation results.
At $n_e\ll n_c$, the magnetic loop strength is approximated as $B_\varphi\propto a_0n_e$ by using the balance between magnetic pressure $P_B=B_\varphi^2/(8\pi)$ and electron thermal pressure $P_e=\varepsilon_en_e$ [see Fig.~\ref{fig:fig5_para_scan}(b)], where the electron energy is estimated as $\varepsilon_e\propto E_b E_aJ_0(k_br)\sim a_0^2n_e$ [see Fig.~\ref{fig:fig5_para_scan}(c)] based on the increase by the coupled term in Eq.\eqref{eq:dgg_dt}. 
At $n_e\gtrsim 0.1n_c$, however, the loop strength $B_\varphi\propto a_0$ does not increase with the rise of $n_e$ since the increased energy transferred to plasma electrons compensates the increment of the EM energy converted to loops. Using $P_e\sim P_B$, one obtains $\varepsilon_e\propto a_0^2/n_e$ where the electron energy $\varepsilon_e$ decreases with the increase of $n_e$ [see Fig.~\ref{fig:fig5_para_scan}(c)]. 
%
%
Given that transverse filamentation and scattered EM waves are prerequisites for the occurrence of the magnetic loop, the criterion is estimated as $n_e\gtrsim n_e^* \sim \mathrm{max}\left\{(a_0\omega_0 t)^{-1}n_c, (\omega_0 t)^{-1}n_c\right\}$ [see Fig.~\ref{fig:fig5_para_scan}(d)].
%

It should be noted that the results presented in this work are primarily based on PIC simulations employing lateral periodic boundary conditions. This configuration is widely adopted in long-pulse, non-relativistic laser-plasma interaction studies to mitigate substantial computational costs~\cite{gu2021multi,liu2023parametric,rovere2026effects}. To verify the robustness of the observed light scattering processes, we performed an additional simulation using open boundary conditions with a transverse laser spot size of $\sigma_0=100\mu m$, while keeping all other parameters identical. Nevertheless, a more systematic investigation across a broader parameter space is warranted in future studies to fully quantify the potential influence of periodic boundaries.

In summary, we have identified a potent electron heating mechanism, magnetic loop penetration, which arises from the nonlinear evolution of picosecond, nonrelativistic laser-pulses in under-critical plasmas. Through large-scale 2D-PIC simulations using the EPOCH code, we have demonstrated that the interaction is governed by a rapid transition from stimulated scattering and filamentation  to a robust state of kinetic turbulence. This turbulent environment is characterized by the formation of self-consistent magnetic loops associated with electron density cavities, which are analogous to post solitons sustained in relativistic laser-driven plasma channel.
Our analysis reveals that these magnetic structures act as localized acceleration sites, where the dissipation of trapped EM energy and barrier penetration drive the production of a high-energy electron tail beyond MeV scales, which is orders of magnitude higher than the ponderomotive scaling. This mechanism provides a self-consistent kinetic explanation for the anomalous electron heating observed in ignition-scale experiments~\cite{campbell2017laser,meezan2017indirect,scott2021shock,batani2023future,he2015physical,zhang2020enhanced}, which cannot be fully accounted for by linear wave-breaking theories~\cite{estabrook1980heating,baldis1991coexistence,rousseaux1992suprathermal}. Given the ubiquity of long-pulse propagation in the corona of ICF targets and the expanding capabilities of multi-kilojoule laser systems, the magnetic loop penetration heating identified here represents an inevitable pathway for energy partition in HEDP plasmas.

\begin{acknowledgments}
This work was partially supported by NSF Grant PHY-2308641. The PIC code EPOCH is funded by the UK EPSRC grants EP/G054950/1, EP/G056803/1, EP/G055165/1 and EP/ M022463/1. 
This work used Delta-cpu at the National Center for Supercomputing Applications (NCSA) through the allocation of PHY230120 and PHY230121 from the Advanced Cyberinfrastructure Coordination Ecosystem: Services \& Support (ACCESS) program, which is supported by National Science Foundation grants \#2138259, \#2138286, \#2138307, \#2137603, and \#2138296.
Z.G. acknowledges the HPC Cluster of ITP-CAS for providing computational resources and the CAS Project for Young Scientists in Basic Research (Grant No. YSBR-141).
\end{acknowledgments}

\appendix

\section{\NoCaseChange{Appendix A: The parameter setup in PIC simulations}}\label{Appendix_A}
The detailed parameter setup in PIC simulations can be found in Table~\ref{tbl:params} below. Additionally, in 2D PIC simulations of plasma current filamentation instabilities [shown in Fig.~\ref{fig:fig4_e_acc}(b)], the domain in $(x,y)$ space has a dimension of $20\mathrm{\mu m}\,\times\,20\mathrm{\mu m}$, with a cell size of $\Delta x=\Delta y=1/40\,\mathrm{\mu m}$. An electron beam with density $n_{be}=3.6\times10^{18}\,\mathrm{cm}^{-3}$ and energy $\gamma_0=5$ is initialized to propagate along the $+z$ direction. A background plasma electron flow with a density $n_{pe}=5.5\times10^{19}\,\mathrm{cm}^{-3}$ and velocity $v_z\sim (n_{be}/n_{pe})c$ ($c$ is the speed of light) is set to neutralize the current density at the initial time.
The ions, with charge $Z_i=1$, mass $m_i=1836m_e$, and density $n_{pi}=n_{be}+n_{pe}$, are stationary to neutralize the charge density at the beginning. The computational area is filled with 20 macro-particles per cell for each species. For comparison, the energy of the incident electron beam is roughly similar to that of the incident EM wave in the main simulation of electron penetration acceleration.

\begin{figure*}[t]
\centering
\includegraphics[width=1.3\columnwidth]{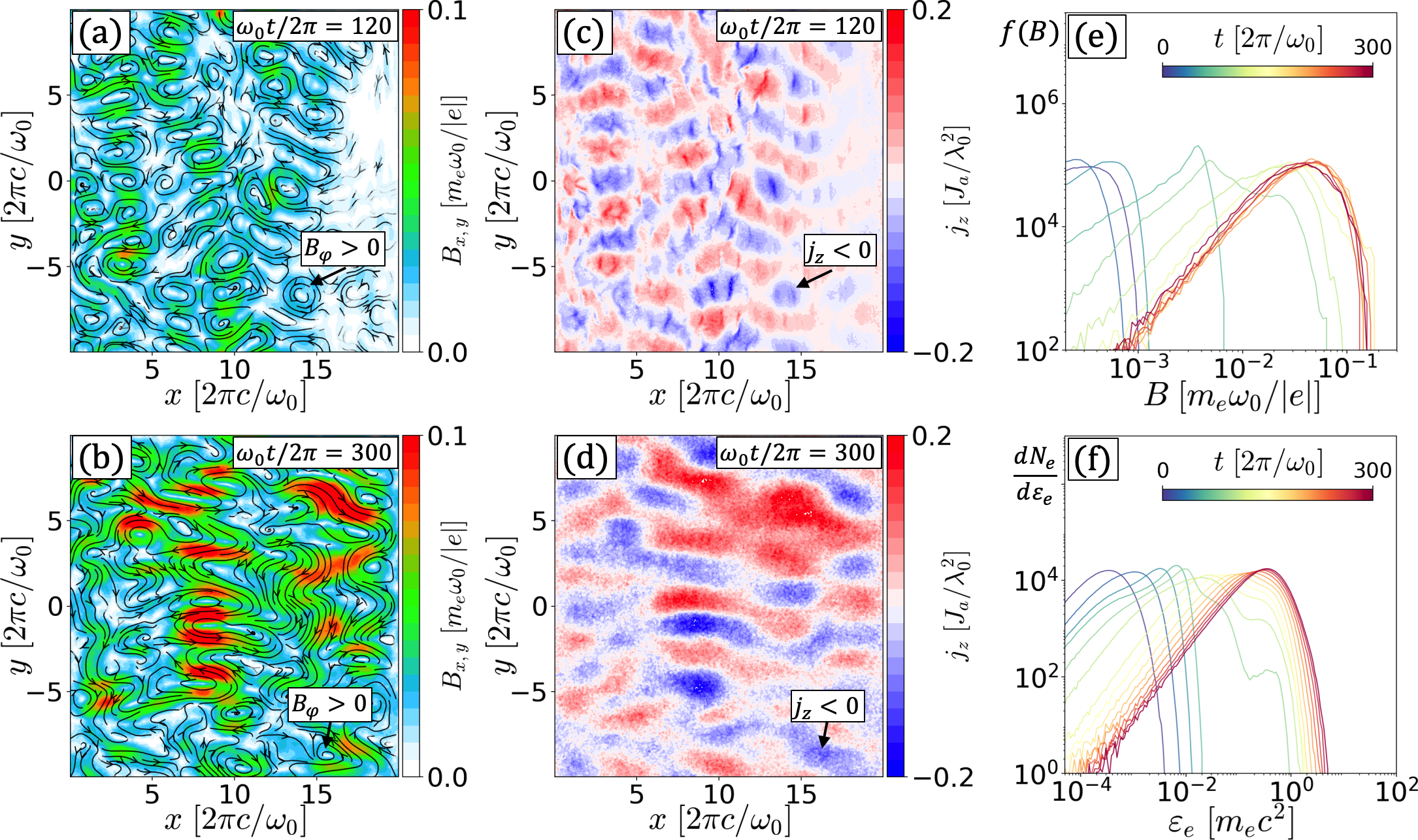}
\caption{3D PIC simulation results. (a)(b) and (c)(d) present the spatial distribution of the the magnetic field $B_{xy}$ and the and the current density $j_z$ at the slice of $z=0$ for the time $\omega_0 t/(2\pi)$ = 120 and 300, respectively. The black streamlines in (a)(b) denote the magnetic field direction.
(e) shows the time-evolved magnetic spectra $f(B)$ vs $B$. 
(f) shows the electron energy spectra $dN_e/d\varepsilon_e$ vs $\varepsilon_e$.}
\label{fig:res_sim_3d}
\end{figure*}

In addition, 3D PIC simulation results qualitatively reproduced the feature of electron penetration acceleration in turbulent magnetic loops shown by 2D PIC simulations.
The 3D simulation is calculated within a domain of $20\mathrm{\mu m}\times 20\mathrm{\mu m}\times 20\mathrm{\mu m}$ with a grid resolution of $1/20\times1/20\times 1/20\,\mathrm{\mu m}^{3}$.
A linearly polarized laser pulse (with the electric field along the $z$-direction) was incident from the left boundary.
The laser pulse has a peak intensity of $I_0=5.5 \times 10^{16}\mathrm{W/cm}^2$, equivalent to the normalized amplitude $a_0\equiv eE_l/(m_ec\omega_0)= 0.2$ with the wavelength $\lambda_0=2\pi c/\omega_0=1\mu m$ and $c$ the speed of light. 
The pulse was transversely infinite and had $1\,$ps duration.
The plasma electron and proton densities were $n_e=n_i=0.1n_c\approx 1.1\times10^{20}\mathrm{cm}^{-3}$ and the proton mass is $m_i=1836m_e$. The temperature is $T_{e,i}=10\,$eV for both species.
A periodic boundary condition was used for the lateral sides and an open condition was used for the longitudinal boundaries. 

The distributions of $B_{xy}$ and $j_z$ in Fig.~\ref{fig:res_sim_3d}(a)(b)(c)(d) show the typical magnetic loop structures and the localized current density $j_z$. Here, the magnetic loops have the left-hand chirality with $(\nabla\times\hat{B}_\varphi)\cdot j_z<0$.  
The time-evolved magnetic spectra $f(B)$ vs $B$ in Fig.~\ref{fig:res_sim_3d}(e) indicate that the formation of magnetic loops begins after $\omega_0t/2\pi>100$ and the turbulent loops become the dominant structure at $\omega_0t/2\pi\sim 300$. The time-evolved electron spectra $dN_e/d\varepsilon_e$ vs $\varepsilon_e$ in Fig.~\ref{fig:res_sim_3d}(e) illustrate that a considerable fraction of electrons can be accelerated to an energy beyond MeV.  
Although the growth of magnetic loops becomes faster in the 3D scenario compared to the 2D case, the results shown in Fig.~\ref{fig:res_sim_3d} confirm that the 3D PIC simulation qualitatively reproduced the 2D PIC simulation.

Besides, the 3D magnetic structure is indeed more complicated than the 2D case. The difference in the simulation results comes from the EM eigenmode trapped in the density cavity having a cylindrical symmetry in the 2D situation, but this is not realistic in the 3D geometry with a finite-scale size along the $z$-direction. Additionally, in 3D scenarios, depletion of the transverse laser field $E_z$ would lead to a local current $j_z$ associated with the occurrence of electron vortex~\cite{lezhnin2018annihilation} or magnetic island structures~\cite{gong2021retrieving}. These effects would also complicate the magnetic field structure in the 3D cases, which is worth further studies in the future work.


\section{\NoCaseChange{Appendix B: Eigenstate of magnetic loops}}\label{Appendix_B}
The wave equation in the cylindrically symmetric form is written as
\begin{align}
\frac{1}{r} \frac{\partial}{\partial r} \left(r \frac{\partial u}{\partial r}\right) + \frac{1}{r^2} \frac{\partial^2 u}{\partial \varphi^2} + \frac{\partial^2 u}{\partial z^2} - \frac{1}{c^2} \frac{\partial^2 u}{\partial t^2} = 0.
\end{align}
The electric field along the $z$-axis can be characterized as 
\begin{align}
    E_z(r,\varphi,z,t)=E_0 f(r,\varphi)\cos (k_zz)e^{i\omega t}
\end{align}
with $k_z^2=\frac{n_z^2\pi^2}{L_z^2}$.
Then the wave equation becomes
\begin{align}
\frac{1}{r} \frac{\partial}{\partial r} \left(r \frac{\partial f}{\partial r}\right) + \frac{1}{r^2} \frac{\partial^2 f}{\partial \varphi^2} + k_b^2 f = 0
\end{align}
with $k_b^2=\frac{\omega^2}{c^2}-k_z^2$.
Considering the variable separation by $f(r,\varphi)=g(r)e^{im\varphi}$, one can rewrite the above formula as the Bessel equation
\begin{align}
    \mathcal{R}^2\frac{d^2g}{d\mathcal{R}^2}+\mathcal{R}\frac{dg}{d\mathcal{R}}+(\mathcal{R}^2-m^2)g=0
\end{align}
with $\mathcal{R}=k_br$.
The independent solutions of the equation are $ J_m(\mathcal{R})$ and $ Y_m(\mathcal{R})$. The $ J_m(\mathcal{R})$ are regular at $\mathcal{R}=0$ , whereas the $ Y_m(\mathcal{R})$ are singular. The radial eigenfunction must be regular at $ r=0$, which rules out the $ Y_m(\mathcal{R})$ solutions. 
Thus, the most general solution is
\begin{align}
    \displaystyle E_z(r,\varphi,z,t) = E_0\, J_m(k_b\, r) \,{\rm e}^{\,{\rm i}\,m\,\varphi} {\rm e}^{i\omega t}.
\end{align}
where $ \cos(k_z z)=1$ is used for $L_z\rightarrow\infty$ in the 2D case.
$k_b$ is determined by the solution of $J_m(k_b r_b) = 0$, where $r_b$ is the radius of the loop. 

The corresponding $B_\varphi$ is derived by $\partial B_\varphi/\partial t=\partial E_z/\partial r-\partial E_z/\partial z$ as
\begin{align}
    \displaystyle B_\varphi(r,\varphi,t) =\frac{i k_b}{\omega}E_0\, J_1(k_b\, r) {\rm e}^{i\omega t}.
\end{align}
Specifically for $m=0$,
\begin{align}
    \displaystyle B_\varphi(r,\varphi,t) = \frac{i k_b}{\omega}E_0\, J_1(k_b\, r) {\rm e}^{i\omega t},
\end{align}
where $J_0'(k_br)=d J_0(k_br)/dr=-k_bJ_{1}(k_br)$. 
Using $r=r_b$ at the edge and $J_0(k_b r_b)=0$, one can obtain $k_b r_b=2.405$ and $\omega\approx 2.41c/r_b$.

\section{\NoCaseChange{Appendix C: Electron dynamics in the combined field}}\label{Appendix_C}

The electron dynamics are governed by the Lorentz equation
$
\frac{d\bm{p}}{dt}=-|e|(\bm{E}+\bm{v}\times\bm{B}),
$
which is reformulated as 
$
\frac{dp_x}{dt}=-|e|(E_x+v_yB_z-v_zB_y),
\frac{dp_y}{dt}=-|e|(E_y+v_zB_x-v_xB_z),
\frac{dp_z}{dt}=-|e|(E_z+v_xB_y-v_yB_x)=|e|\frac{dA_z}{dt},
$
where $\bm{E}=-\partial \bm{A}/\partial t$ and $\bm{B}=\nabla\times\bm{A}$.
Considering the background electromagnetic wave (i.e. laser wave) $E_{a,z}=E_a e^{i(\omega_0 t-k_0 x+\phi_{a0})}$ with $B_{a,y}=-(E_a/c)e^{i(\omega_0 t-k_0 x+\phi_{a0})}$ and the magnetic loop field $B_{b,\phi}=(E_b/c)J_1(k_b r) e^{i(\omega_b t+\pi/2+\phi_{b0})}$ with $E_{b,z}=E_bJ_0(k_b r) e^{i(\omega_b t+\phi_{b0})}$, the corresponding field vector potentials are expressed as
\begin{eqnarray}
&& A_{a,z}=i\frac{E_a}{\omega_0}e^{i(\omega_0 t-k_0 x+\phi_{a0})},\\ 
&& A_{b,z}=i\frac{E_b}{\omega_b}J_0(k_b r)e^{i(\omega_b t+\phi_{b0})}.
\end{eqnarray}
Here, $r=\sqrt{(x-x_b)^2+(y-y_b)^2}$ and $(x_b,y_b)$ is the center of the magnetic loop. 
The electron momentum along the $z$-axis is
$
p_{z}=A_{a,z}+A_{b,z}
$.
Then, the electron energy increment is formulated as
\begin{eqnarray}
\frac{d\varepsilon_e}{dt}\sim -\frac{|e| p_z E_z}{\gamma}\sim -|e|p_zE_z\ \  \mathrm{at}\ \ \gamma\sim 1,
\end{eqnarray}
with
\begin{widetext}
\begin{eqnarray}
 -p_zE_z &&= -\left[\frac{E_a}{\omega_0}\cos(\omega_0 t-k_0 x+\frac{\pi}{2}+\phi_{a0}) + \frac{E_b}{\omega_b}J_0(k_b r)\cos(\omega_b t+\frac{\pi}{2}+\phi_{b0})\right] \nonumber\\
 &&\ \ \ \ \ \cdot\left[E_a\cos(\omega_0t-k_0x+\phi_{a0}) + E_bJ_0(k_b r)\cos(\omega_b t+\phi_{b0})\right]  \nonumber\\
  &&= \frac{E_a^2}{2\omega_0}\sin[2(\omega_0 t-k_0x+\phi_{a0})] + \frac{E_b^2J_0^2(k_br)}{2\omega_b}\sin[2(\omega_b t+\phi_{b0})] \nonumber\\
 &&\ \ \ \ +\frac{E_aE_bJ_0(k_br)}{2\omega_0\omega_b} [(\omega_b+\omega_0)\sin(\omega_0t-k_0x+\phi_{a0}+\omega_bt+\phi_{b0}) \nonumber\\ 
 &&\ \ \ \ +(\omega_b-\omega_0)\sin(\omega_0t-k_0x+\phi_{a0}-\omega_bt-\phi_{b0})]
\end{eqnarray}
\end{widetext}

%

\end{document}